\documentclass[namedreferences,hyperref,optionalrh]{spr-sola}
\usepackage{lmodern}
\usepackage{graphicx}        
\usepackage{color}           
\usepackage{soul}
\usepackage{todonotes}
\usepackage{amsmath}

\chardef\us=`\_

\def\ers{ERs}

\newcommand{\anu}[1]{\textcolor{black}{#1}}

\begin{document}

\begin{frontmatter}
\title{Solar Ephemeral Regions: Hale's and Joy's laws}

\author[addressref=aff1,email={anubsreedevi.rs.phy20@itbhu.ac.in}]{\inits{A.}\fnm{Anu}~\snm{Sreedevi}\orcid{0000-0001-7036-2902}}
\author[addressref=aff2,email={}]{\inits{B.K.}\fnm{Bibhuti~Kumar}~\snm{Jha}\orcid{0000-0003-3191-4625}}
\author[addressref={aff1}, corref,email={karak.phy@iitbhu.ac.in}]{\inits{B.B.}\fnm{Bidya~Binay}~\snm{Karak}\orcid{0000-0002-8883-3562}}
\author[addressref={aff3}, corref,email={ }]{\inits{R}\fnm{Robert}~\snm{Cameron}\orcid{ }}

\address[id=aff1]{Department of Physics, Indian Institute of Technology (Banaras Hindu University), Varanasi 221005, India}
\address[id=aff2]{Southwest Research Institute, Boulder, CO 80302, USA}
\address[id=aff3]{Max Planck Institute for Solar System Research, Justus von Liebig Weg, 37077 Göttingen, Germany}

\runningauthor{Sreedevi et al.}
\runningtitle{Solar Ephemeral Regions: Hale's and Joy's laws}

\begin{abstract}
Besides the active regions and sunspots, the Sun produces many bipolar structures 
such as 
ephemeral regions (ERs) and even smaller magnetic 
bipoles. 
In this study, we investigate the smallest end of bipolar magnetic structures on the Sun, focusing on ERs using  Helioseismic and Magnetic Imager (HMI) full-disk LOS magnetograms. We identify and track the emergence–decay sequences of ERs to quantify their flux and morphological properties. 
Consistent with previous studies, we find that the flux distribution follows a log-normal profile and the total unsigned flux emergence rate is estimated to be around $1.45 \times 10^{22}$\,Mx/day.
We find the ERs 
are 
almost uniformly distributed
on the solar disc with a substantial fraction
emerging outside the
activity belts. A statistical analysis of ER's tilt angle reveals a weak adherence to Hale's polarity law. 
The mixed signatures suggest that a majority of the ERs are associated with the large-scale global dynamo, while the remainder most likely originate from the small-scale dynamo. Overall, our findings indicate that even the small bipolar magnetic structures are statistically imprinted by the Sun's global dynamo. Our results bridge the scale between ERs and active regions, providing valuable constraints on how magnetic flux is structured and emerges across various spatial and temporal scales on the Sun.
\end{abstract}
\keywords{Active Regions, Sunspots, Magnetic fields, Solar Cycle, Observations}
\end{frontmatter}

\section{Introduction}
     \label{Intro} 

Darker features usually observed on the visible surface of the Sun are commonly known as sunspots \citep[for a comprehensive \anu{review on sunspots}, see][and references therein]{S2003}. They typically appear in pairs or groups, and the corresponding line-of-sight (LOS) magnetic field measurements reveal opposite magnetic polarities of strong, organized fields, collectively referred to as Bipolar Magnetic Regions (BMRs; \citealt{H1991, Howard96}). Sunspots are the most extensively studied solar features with records that span several centuries, and they are the prominent and easily observable manifestations of the dynamo operating beneath the surface \citep{H08, SG1999, HA2015}. The solar dynamo operates by cyclically converting the large-scale poloidal magnetic field into a toroidal configuration, whose buoyant flux bundles rise through the convection zone to produce the characteristic bipolar structures of BMRs \citep{P1955}. These BMRs are typically observed to be tilted with respect to the solar equator, and statistically, it is found in the datasets that the tilt increases with latitude -- Joy’s law \citep{H08, H1991, WS1991, AV2020}. Furthermore, the subsequent decay and dispersal of the BMR flux contributes to the cancellation of existing and buildup of new poloidal field (Babcock--Leighton mechanism), thus sustaining the solar magnetic cycle \citep{B1961, L1964, CS2023}.  

A long-lived sunspot harbors a total unsigned magnetic flux roughly on the order of $10^{22}$ Mx or more \citep{WW1995}. A closer inspection of solar magnetograms reveals the presence of coherent bipolar magnetic structures across a wide range of spatial and flux scales, including regions with flux much lower than that of sunspots. Not all of these regions produce sufficient (or not at all) intensity contrast to appear as classical sunspots in white-light observations \citep{TP2014, JK2020}. However, those with total unsigned flux in the range of $10^{20}$–$10^{22}$ Mx still represent and show classical BMR traits. Such intermediate-scale regions follow the solar cycle variation and exhibit the well-known properties of sunspots, and considered to be the manifestation of the global/large-scale solar dynamo \citep{SJ2023, SJ2024, sreedevi25b}.   

At even smaller scales—typically with total unsigned flux between $10^{18}$\,--\,$10^{20}$\,Mx, we find the emergence of coherent bipolar magnetic features known as ephemeral regions \citep[ERs;][]{HH1990, HA1993}. Although they do not produce visible sunspots, ERs share the fundamental bipolar nature of larger magnetic regions. They generally have lifetimes of a few hours to a few days \citep{H2001}, with emergence rates far exceeding those of typical BMRs, and they appear across the solar surface, from quiet regions to the vicinity of large active regions \citep{HA1993, LG2015, HD2008, TV2010}. By combining high-resolution measurements of small and larger BMRs, \citet{TP2011} found that the flux emergence rate of ERs and BMRs together follows a single power-law distribution spanning nearly seven orders of magnitude in total flux \citep[$10^{16}$–$10^{23}$ Mx; also see][for the distribution in the low-flux range]{Smitha17}. ERs were also found to exhibit cyclic behavior similar to BMRs, though their cycles tend to be more extended in duration and of lower amplitude, contributing to the extended minima observed in solar activity cycles \citep{BM1992}. Their irregularity in emergence, orientation and evolution distinguish them from conventional BMRs, yet they contribute significantly to the global magnetic flux budget and play a non-negligible role in the operation of the solar dynamo \citep{HH1990, HK2024}.  

In the present study, we focus on the lower end of ER flux distribution, aiming to bridge the scale between ERs and BMRs. Our primary objective is to determine whether these small-scale bipolar features exhibit trends similar to those observed in BMRs, particularly with regard to their adherence to Hale's polarity law and their tilt orientation.
The paper is organized as follows: Section 2 describes the datasets and the detection and tracking methods employed, while Section 3 presents the results and analysis. Finally, Section 4 summarizes our findings and discusses their implications.

\section{Data and Methodology}

\subsection{Data}
To study ERs we use line-of-sight (LOS) magnetograms obtained from the Helioseismic and Magnetic Imager (HMI) onboard the Solar Dynamics Observatory \citep[SDO;][]{SS2012, SchouS2012}. HMI provides full-disk solar observations with a spatial resolution of approximately 1\,arcsec (0.5\,arcsec/pixel) with a cadence of $45$\,seconds and 720\,seconds. The instrument measures the Fe {\sc I} 6173\,\AA\ absorption line to infer the photospheric magnetic field with high sensitivity and stability. For the purpose of this study, LOS magnetograms series, `hmi.M\_720s' with 720\,seconds cadence is used. Given that ERs typically evolve over timescales of several hours \citep{ TV2010, TP2011}, the 720\,seconds (12-minutes) cadence is sufficient to capture their evolution while avoiding the heavy computational load associated with the high cadence data set.

\subsection{Automated Detection and Pairing of ERs}

The detection part of the algorithm is \anu{inspired by} the ER detection procedure from \citet{TV2010} with modifications.  For the analysis, we consider the area within $0.8 \,\mathrm{R}_\odot$ from the disk center to avoid severe projection and limb-related artifacts. We begin by applying a Gaussian smoothing with a FWHM of 12\,pixels to suppress granular noise while preserving the spatial integrity of  ephemeral structure. This scale was found optimal for maintaining ephemeral-scale magnetic coherence without introducing incorrect pairings. Further, we correct for projection effects in the magnetograms, converting the LOS field to the radial component by dividing by the cosine of the heliocentric angle.

Coherent positive and negative magnetic structures were identified by applying a threshold of $\pm30$\,G on the magnetogram. This value lies above the typical noise level of HMI LOS measurements ($\approx10$\,G) and efficiently isolates magnetic flux concentrations, while excluding weak background salt and pepper fields. To ensure compactness and continuity of the detected regions, we apply a morphological dilation to the thresholded regions using a circular kernel of radius 4\,pixels. This results in isolated magnetic islands with sharp boundaries. Visual inspection of the detected maps shows that structures smaller than 100\,pixels are generally highly irregular and appear as isolated noise fragments rather than stable bipolar features. These are therefore excluded from further analysis. Similarly, coherent structures larger than 2000\,pixels, corresponding to parts of active regions are rejected to exclude large magnetic concentrations, thereby retaining only magnetic features of ephemeral scales. For each identified region, the centroid location (both in heliographic latitude and longitude) and total magnetic flux were calculated, along with the maximum field strength within the region. These quantities provide a basis for magnetic feature characterization and subsequent pairing.

Candidate bipolar magnetic regions were identified by pairing positive and negative polarity patches on the basis of spatial proximity and flux balance. For each candidate region, only opposite-polarity regions within a maximum angular separation of $1.5^\circ$ \citep[18\,Mm, the separation previously reported for ephemeral scales;][]{SJ2024, GS2026} were considered, ensuring that the paired regions are close neighbors on the solar surface. Among these nearby candidates, the region whose flux most nearly balances that of the source was selected. The flux balance $r$ was defined as
\begin{equation}
\text{$r$} = \frac{\left| \Phi_\mathrm{+} + \Phi_\mathrm{-} \right|}{\max\left( \left| \Phi_\mathrm{+} \right|, \left| \Phi_\mathrm{-} \right| \right)},
\end{equation}

where, $\Phi_+$ and $\Phi_-$ are the nearest positive and negative flux regions. Only pairs satisfying flux balance $r \leq0.8$ were retained, ensuring that both polarities contribute comparably to the total unsigned flux. We would like to note that, \anu{we detect ERs across all ranges of flux balance between 0.0 and 0.8}, and the details of variation in $r$ in the tracked ERs are discussed in the Section~\ref{sec:flux_dist}. To avoid ambiguous pairings, a reverse check for the opposite pair is performed: a candidate pair is retained only if each region identified the other as its best match. This procedure reliably isolates coherent, flux-balanced bipolar structures while minimizing incorrect pairings due to unrelated magnetic fragments.

\begin{figure}
    \includegraphics[width=\textwidth]{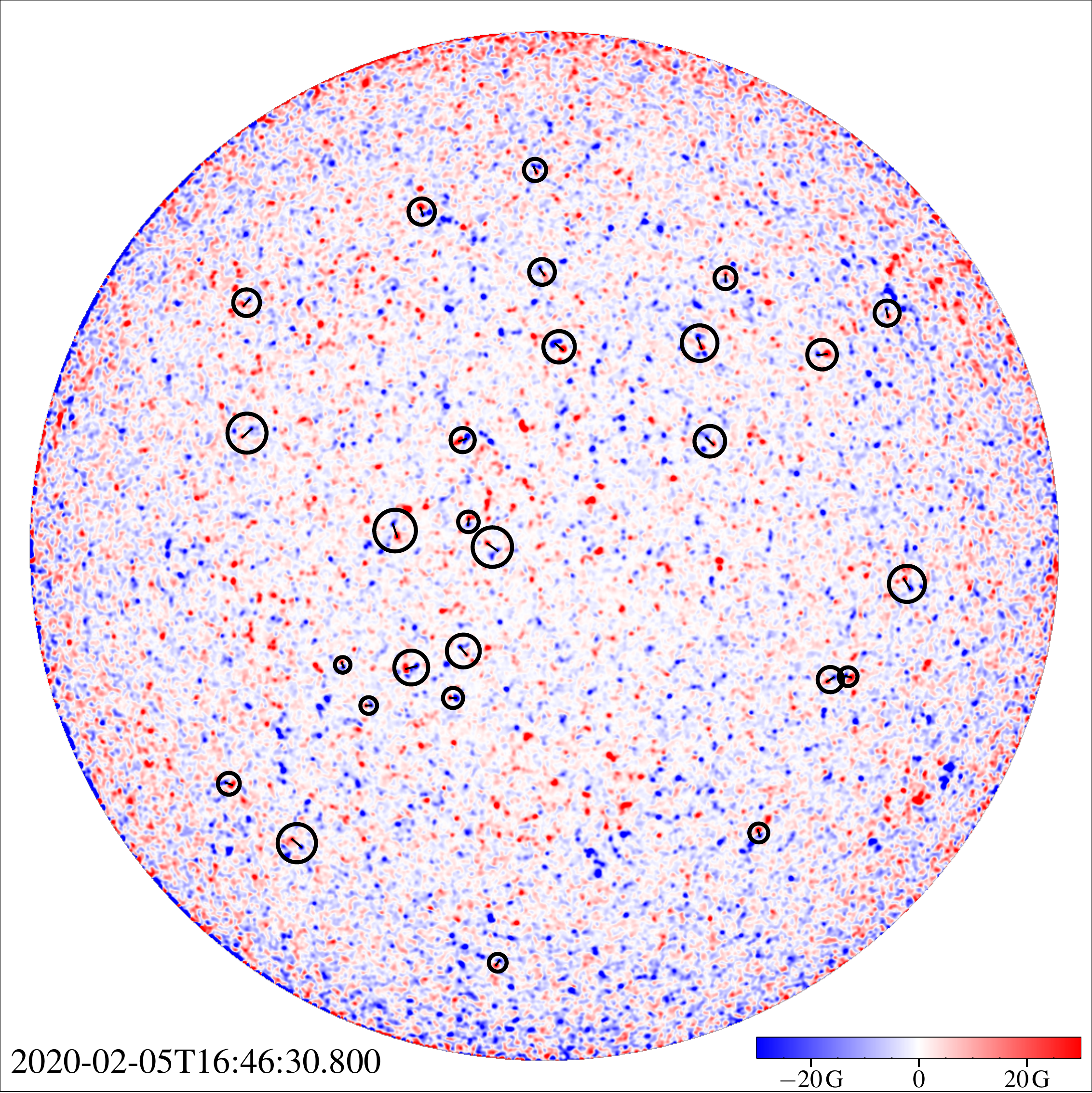}
    \caption{HMI Magnetogram for February 05, 2020 16:46 UTC. The magnetogram is corrected for the projection effects and \anu{a Gaussian smothing with a FWHM of 12 pixels is applied
}. Identified ERs are marked by outlined circles. The color bar is shown at the bottom of the figure. The same color scale is used for all magnetograms throughout this manuscript.}
\label{fig:det}
\end{figure}

For each magnetogram, the date, heliographic coordinates of both polarities, their total fluxes, and flux balance parameters of pairs of ERs were recorded in an output catalog. This procedure ensures consistent and objective identification of ERs across a large time series of solar magnetograms, forming a reliable basis for subsequent analyses of their emergence properties and statistical behavior. An example of the quiet-sun ERs detected for the day February 05, 2020 is shown in Figure~\ref{fig:det}.

In comparison to the bipole-identification method of \citet{TV2010} \citep[also previously adopted by ][]{SP2002} our procedure introduces several important refinements. We restrict measurements to within $0.8\,{\rm R}_\odot$ and correct for projection effects to minimize projection and limb artifacts, whereas \citet{TV2010} rely on size bases thresholding without a geometric filtering. We additionally apply Gaussian smoothing to suppress granular noise while preserving coherent flux structures. 
Our pairing involves a relaxed flux balance  criterion ($\le0.8$) and a stricter polarity separation criterion between the polarities, while \citet{TV2010} allowed larger separations ($\sim 2^\circ$) rather than our $\sim 1.5^\circ$  and a stricter 50\% flux-match tolerance.

\subsection{Tracking of ERs}

Following the identification of ERs from individual magnetograms, we implemented a temporal tracking procedure to follow their evolution across consecutive observations (separated by 12\,minutes in time). This step is very crucial, as we are dealing with flux emergence in ephemeral scales. At such length-scales it becomes difficult to differentiate between small scale \anu{unipolar regions coming together} (background salt and pepper field) to true emerging compact \ers. Previous studies suggest that \ers\ occupy the lower end of bipolar emergence spectrum and are known to survive for few hours to days, unlike the background field that changes more frequently. Hence tracking the \ers\ becomes a unique way to filter out faulty detections and retain true emerging ERs. The tracking was performed on the detected positions of positive and negative polarity patches extracted from the output files generated from the detection procedure. 

\begin{figure}
    \includegraphics[width=\textwidth]{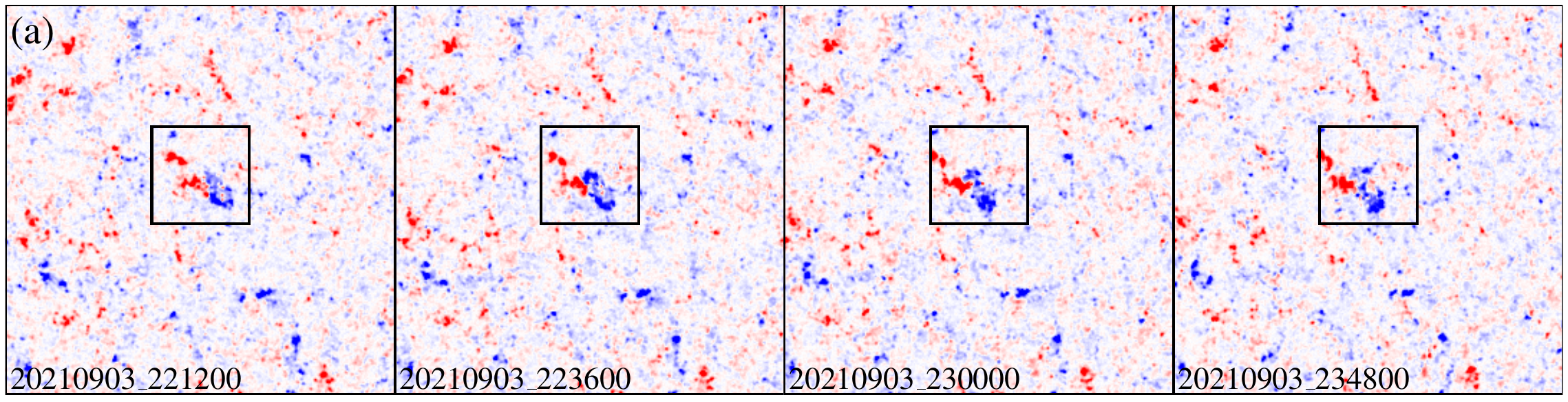}
    \includegraphics[width=\textwidth]{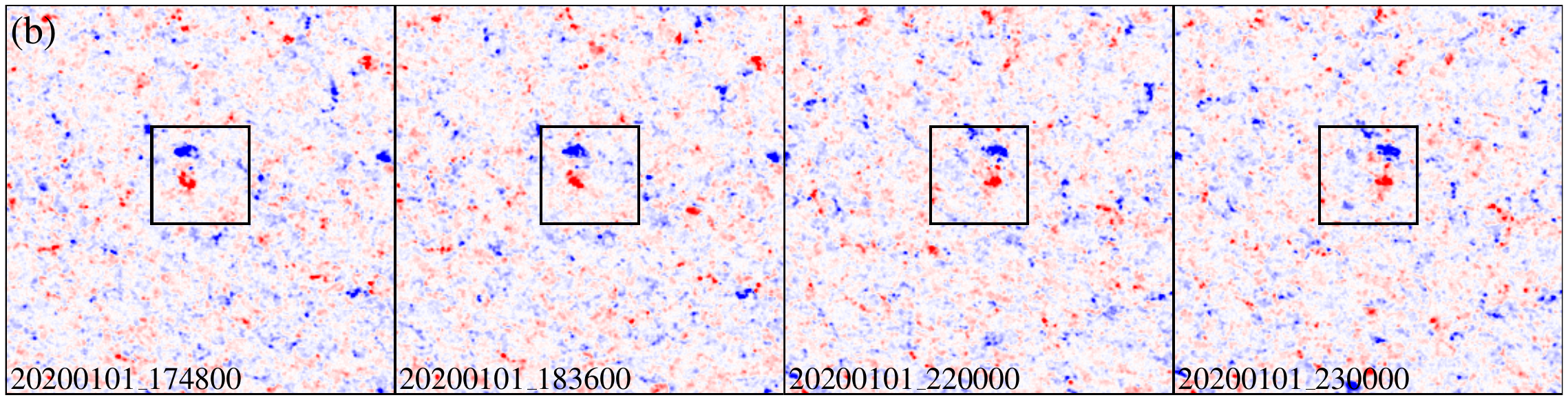}
    \includegraphics[width=\textwidth]{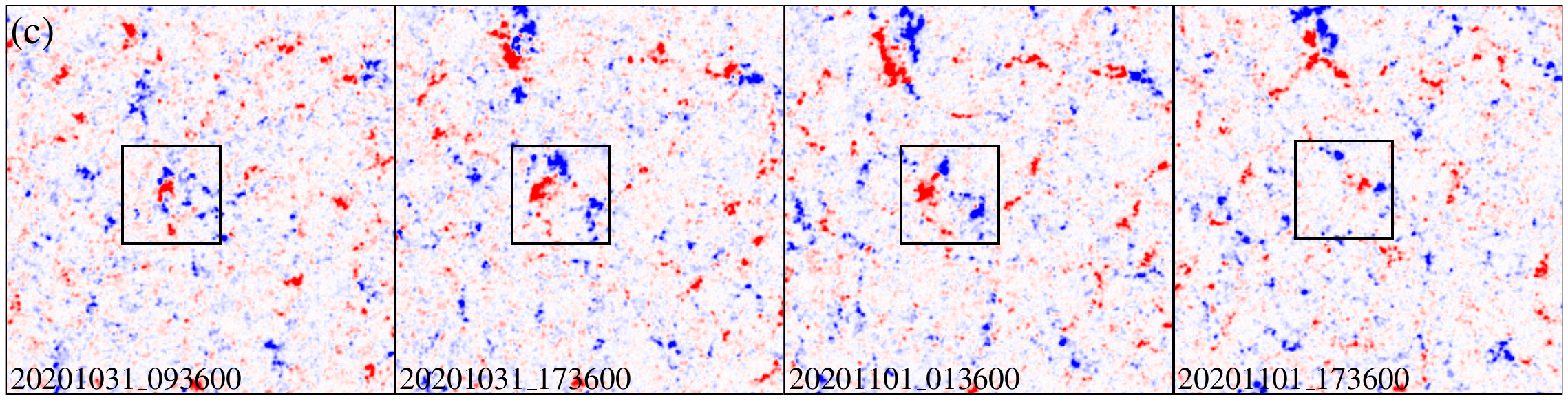}
    \caption{Temporal evolution of three ERs tracked using the automated tracking algorithm. Each row illustrates the development of an individual ER, with successive snapshots separated by roughly one hour. The region of interest is marked by a box, and the date of each snapshot is shown at the bottom of each panel. }
\label{fig:evolution}
\end{figure}

To follow the temporal evolution of detected ERs, each identified ER in a given frame was propagated forward in time to subsequent observations using its heliographic latitude and longitude information. The longitudinal motion of each polarity was corrected for solar differential rotation, adopting a standard latitude-dependent profile \citep{HH1990}: 
\begin{equation}
\Omega(\lambda) = \left[13.45- 3.0 \, \sin^2 \lambda\right]\mathrm{ deg/day},
\end{equation}

where $\lambda$ is the heliographic latitude. The predicted longitude of each polarity in subsequent frames was computed based on this rotation rate and the elapsed time since the last observation. For each predicted ER position, the algorithm searched the subsequent frame for a matching bipolar pair within a maximum tolerance of $1.5^{\circ}$ in latitude and rotation-corrected longitude for both polarities. The \anu{ER's evolution is followed consistently throughout the 5~day period.} 

This forward-prediction approach ensures that each ER is followed consistently over time, providing a reliable record of its emergence, motion, and evolution across multiple frames. Only ERs observed for at least two frames were retained to exclude faulty detections. The final output is a tracked catalog of ERs, including unique identifiers, observation times, and the positions of their positive and negative polarity patches, an example of a tracked ER is shown in Figure \ref{fig:evolution}.

\section{Results}

\subsection{Statistical Properties of ERs}

ERs were systematically identified and tracked over a continuous five-year interval during the ongoing Solar Cycle 25 (January 2020 -- December 2024) using our automated detection and tracking procedures. 
Thus produced catalog consist of \anu{443,022} uniquely tracked ERs. On average, 28 ERs are detected per magnetogram and the magnetic flux associated with each emerging region lies predominantly within the range of $2.73\times 10^{18}$\,Mx to $1.18\times 10^{21}$\,Mx, capturing the classical ER flux distribution. In the coming subsections, we discuss various statistical properties and their solar cycle dependence of the identified ERs on the catalog.

\subsubsection{Flux Distribution of ERs}\label{sec:flux_dist}

\begin{figure}
    \includegraphics[width=0.495\textwidth]{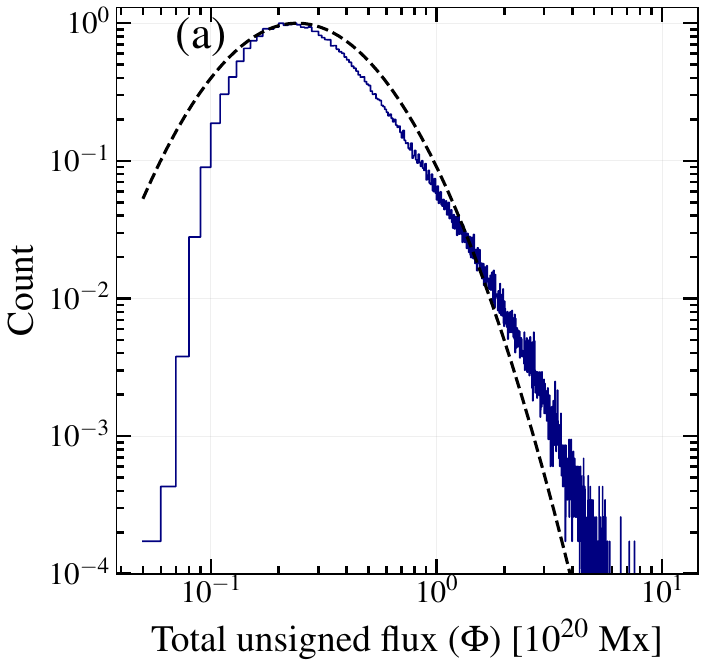}
    \includegraphics[width=0.495\textwidth]{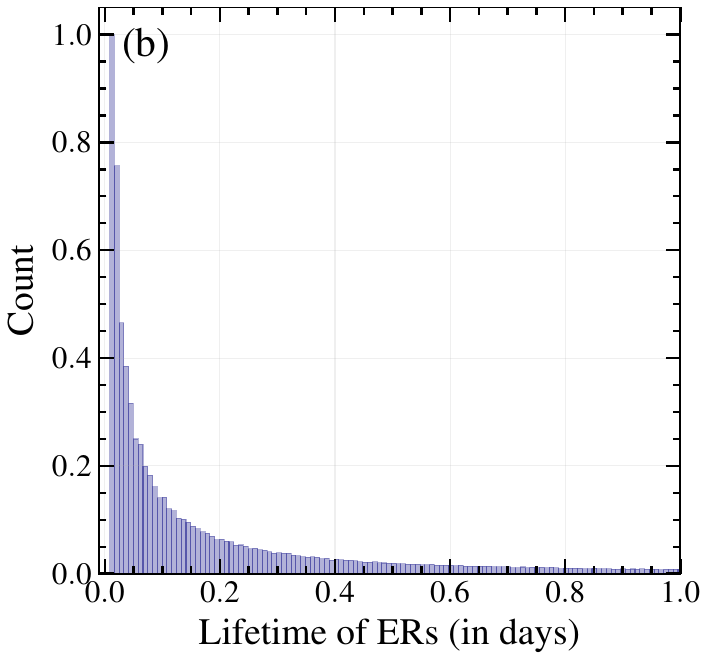}
    \caption{
    (a) The distribution of unsigned magnetic flux of our tracked ERs. For each ER, the representative value corresponds to its peak flux attained during the tracking period. The dashed line represents the fitted log-normal curve. (b) The distribution of tracked lifetime of the ERs.
    }
\label{fig:flux_dist}
\end{figure}

Figure \ref{fig:flux_dist}(a) shows the flux distribution of detected ERs. In the distribution, the representative unsigned total flux of each ER corresponds to the maximum flux attained during its tracked lifetime. It is to be noted that a similar approach is used for other physical parameters, such as heliographic latitude, \anu{Carrington} longitude and the tilt of the ERs. The derived flux distribution, exhibits the characteristic log-normal behavior (with a mean of $4.53 \times 10^{19}$\,Mx), as commonly observed in the fluxes of active regions. The distribution spans more than three orders of magnitude in flux ($10^{18}-10^{21}$\,Mx) and exhibits no distinct break around $10^{20}-10^{21}$\, Mx, suggesting that flux emergence on the Sun is a continous distribution and \anu{ERs do not} form of distinct class, instead stronger BMRs are simply the high-flux tail of the same log-normal ensemble.

The regions in our catalog have a mean total unsigned magnetic flux of $4.56 \times 10^{19}$\,Mx and a median total unsigned magnetic flux of $3.39 \times 10^{19}$\,Mx, which agrees with expectations for the small-scale transient magnetic emergence events on the solar surface with the flux emergence rate of $1.45 \times 10^{22}$ Mx/day, value that is comparable with flux in the active regions. These results highlight the significant contribution of ERs to the Sun’s total magnetic flux budget.

Additionally, we investigate whether the spatial scale of ERs depends on their magnetic flux. For each ER, we use the maximum unsigned flux and measure the polarity separation at the corresponding time. Unlike larger BMRs, where the polarity separation increases with flux, we find no clear dependence of the ER polarity separation on flux, suggesting that the spatial scale of ERs is largely independent of their magnetic flux. 

\begin{figure}
    \centering
    \includegraphics[width=0.495\textwidth]{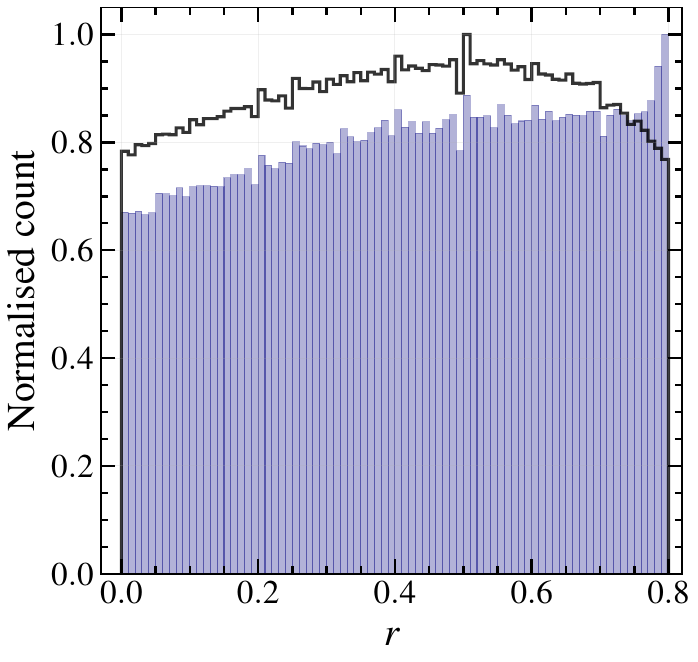}
    \caption{Distribution of the flux imbalance ($r$) of the tracked ERs. The bar chart shows the representative value of $r$ measured when each tracked ER attains its maximum flux during its lifetime, while the stair plot in the front represents the distribution of $r$ for all ER detections at every time step, without considering the tracking. The distribution peaks around $r \approx 0.5$, indicating that most ERs exhibit a moderate degree of flux imbalance between the two polarities.
    }
\label{fig:r_dist}
\end{figure}

It is also noteworthy that the flux imbalance parameter ($r$) of the detected ERs spans a wide range, from nearly perfectly flux-balanced bipoles to the adopted threshold for weakly flux-balanced regions. The distribution of $r$ is shown in Figure~\ref{fig:r_dist}. The bar chart represents the distribution of the representative value of $r$ when an ER attains its maximum flux during the course of tracking, whereas the stair plot represents the distribution of $r$ for all ER detections at every time step, without considering the tracking. As expected, perfectly flux-balanced ERs are comparatively rare. The distribution steadily rises from minimum $r$ and peaks around $r \approx 0.5$, suggesting that the majority of ERs possess a moderate degree of flux imbalance, with the net flux being approximately half of the stronger polarity flux.

\subsubsection{Lifetime of the ERs}
Figure \ref{fig:flux_dist}(b) shows the lifetime 
(time difference between the first and last detections) 
distribution of the tracked ERs. The lifetimes in our catalog range from 12 minutes (the cadence of the observations) to 5 days. The resulting mean in our statistical sample confirms that ERs are generally short-lived in comparison with larger active regions. The upper limit of 5 days arises from the cutoff imposed in our tracking procedure to maintain the reliability of the tracked regions. The mean lifetime of the ERs is 16 hours (0.67 days). The lifetime distribution follows a familiar lognormal behaviour, similar to that reported for sunspots \citep{R1964}, with long-lived regions being relatively rare. In our catalog, only about 20\% of the detected ERs survive for more than one day, those ERs have the mean flux of $6.49 \times 10^{19}$\,Mx.

\subsubsection{Solar Cycle Variation of ERs}

\begin{figure}
    \includegraphics[width=\textwidth]{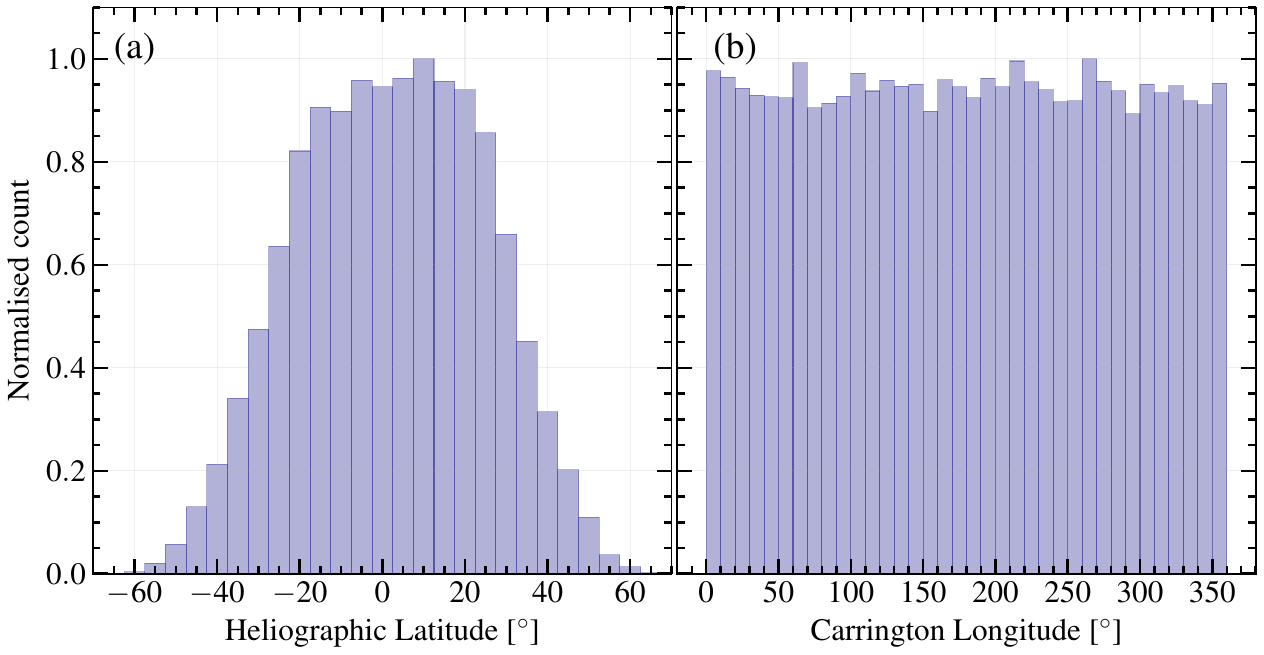}
    \caption{Distributions of the detected ERs in (a) heliographic latitude and (b) Carrington longitude.}
\label{fig:latlon_dist}
\end{figure}

We saw in Section~\ref{sec:flux_dist} that ERs most likely occupy the lower end of the magnetic flux emergence spectrum on the Sun. This naturally leads to the question of whether they also exhibit the characteristic properties of BMRs.
The important characteristics of sunspots is that their number waxes and wanes over the course of the solar cycle. They also predominantly emerge at mid--lower latitude bands, and the active latitude bands gradually migrate towards the equator as the cycle progresses. We would therefore like to examine whether ERs exhibit similar behaviour. Unlike sunspots, in our detections, we do not observe a significant increase in the number of ERs in the rising phase of SC-25, this may be attributed to the methodolgy used for detection of ERs. Meanwhile, the latitudinal distribution of ERs in Figure~\ref{fig:latlon_dist}(a) shows a weak preference to lower latitude bands, their overall distribution is considerably broader, with detections extending to much higher latitudes, outside the typical BMR emergence zone. 

The representative latitude and longitude of each tracked ER are taken at the time when \anu{the observed flux of the ER is at its maximum}. This widespread distribution suggests that a significant fraction of ERs may not be direct surface manifestations of the large-scale dynamo, but instead may originate from processes associated with the small-scale dynamo or from the fragmentation, shredding, and tangling of large-scale magnetic fields as they are transported across the solar surface \citep{KB2016}. A similar absence of preferred emergence zones is also observed in longitude. There has been evidence of active nests/longitudes for BMRs, meanwhile ERs display a nearly uniform longitudinal distribution (see Figure~\ref{fig:latlon_dist}(b)). 

\begin{figure}
    \centering
    \includegraphics[width=0.5\textwidth]{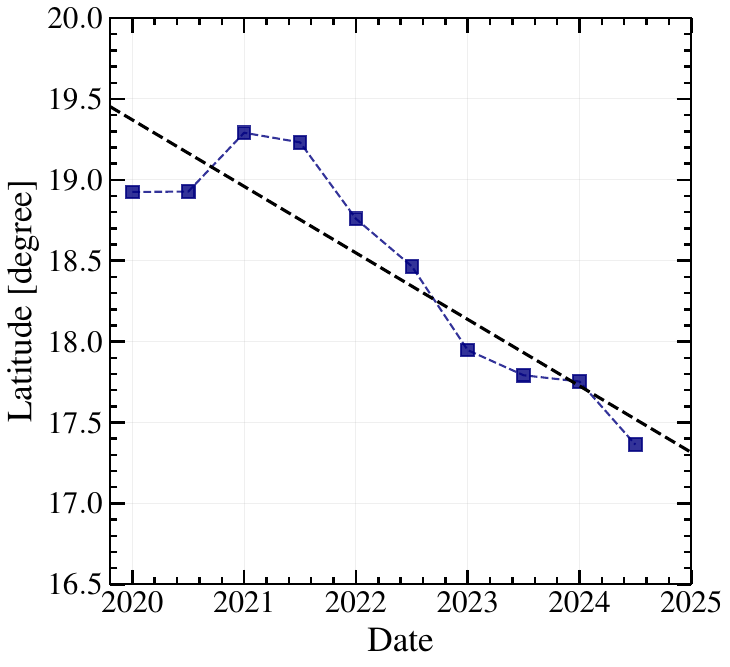}
    \caption{Evolution of the average absolute latitude of ERs during the rising phase of Cycle 25 (2020--2024), computed over successive six-month intervals. \anu{The dashed black line represents the straightline fit with slope of $-0.41$.}}
\label{fig:migration}
\end{figure}

Given that a defining characteristic of the solar cycle is the equatorward migration of the activity belts as the cycle progresses \citep{C1858, M1904}, we investigate whether ERs also exhibit a similar equatorward migration of their active latitude bands, despite their relatively weak preference for specific latitudes. Figure~\ref{fig:migration} shows the evolution of the average absolute latitude of ERs during the initial phase of Cycle 25 (2020--2024), computed over successive two-month intervals. We find a clear and consistent decrease in the mean latitude with time, with a slope of $-0.41$ when fitted with a straightline, indicating an equatorward migration of ER emergence. This behaviour resembles the latitudinal migration observed in sunspots, with a weaker slope, suggesting that the ERs in our catalogue also follow a weak equatorward progression of solar activity over the course of the solar cycle.

\begin{figure}
    \centering
    \includegraphics[width=0.5\textwidth]{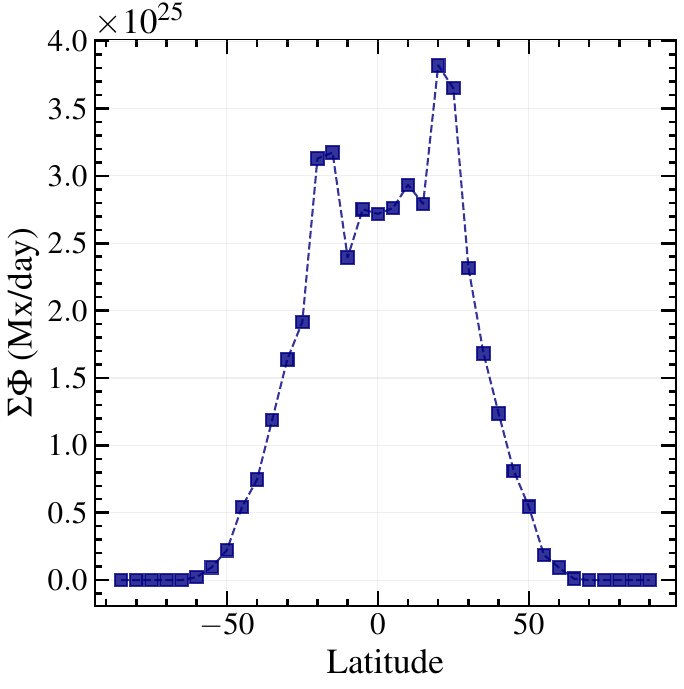}
    \caption{Latitudinal variation of the total flux-emergence rate of ERs, obtained by summing the emerged unsigned flux per unit time within $10^\circ$ latitude bins.}
\label{fig:lat_fluxem}
\end{figure}

In an earlier study, \citet{HH1975} reported a solar-cycle-like dependence in the emergence rate of ERs, with a considerable scatter. In our analysis, such a trend is not clearly evident from the latitudinal distribution. This can be because their study focused on ERs with fluxes around $10^{20}$~Mx, which may largely be associated with \anu{the} large-scale dynamo. Since our sample encompasses a broader range of flux values, including smaller-scale events, this dependence may be diluted or absent in our results. Nevertheless, to quantify the latitudinal dependence \anu{of} emergence rate of ERs, we computed the cumulative flux-emergence rate. Within each $10^\circ$ latitude bin, we calculated the emergence rate by taking all ERs that emerge in that bin, computing their total unsigned flux divided by their tracked lifetimes, and then summing these values in that bin. This summed quantity represents the cumulative flux-emergence rate for that latitude band (Figure \ref{fig:lat_fluxem}). The distribution shows a broad peak at low–mid latitudes, with maximum emergence rates occurring between approximately \anu{$\pm10^\circ-30^\circ$}, followed by a dip at the equator. Although the emergence rate decreases toward higher latitudes, ERs remain detectable even beyond $\pm50^\circ$ reinforcing the picture that ER emergence is not confined to the narrow activity belts characteristic of larger BMRs. This pattern suggests that while ERs preferentially emerge at lower latitudes, their generation mechanism operates over a much wider latitudinal range.

\subsection{Joy's law}

One of the \anu{major characteristics} of bipolar flux emergence on the Sun is its adherence to Hale’s polarity rule and its systematic tilt with respect to the equator. In this section, we examine the degree of order in ER tilt measurements and their adherence to these established patterns.

Any bipolar region observed on the solar surface can have four different orientation classes namely, \anu{Hale--Joy, Hale--anti-Joy, anti-Hale--Joy, and anti-Hale--anti-Joy}, with respect to their obeyance to the Hale's law and Joy's polarity rule. 
We adopt this tilt-angle convention and naming from \citet{JN2021}, where the tilt is defined over the full range [$-180^\circ$, $180^\circ$] (see their Figure 1 for a schematic illustration of these configurations).
We first determine the longitude (with respect to the central meridian) and latitude  center of mass of the positive polarity, denoted by $\phi_+$ and $\lambda_+$ respectively. The location of the negative polarity ($\phi_-$ and $\lambda_-$) is similarly defined. 
The angular distance $d$ between the two polarities is  defined in such a way that $d\in[-180^\circ,180^\circ ]$ and \anu{satisfies}
$
d = \sqrt{\cos^2\left(\frac{\lambda_++\lambda_-}{2}\right) (\phi_+-\phi_-)^2+
(\lambda_+-\lambda_-)^2}
$.
The tilt angle, $\gamma$ is then defined by:

\begin{equation}
\cos\gamma = \frac{\cos\left(\frac{\lambda_+ + \lambda_-}{2}\right) \left(\phi_+-\phi_-\right)}{d}
\quad
\text{and}
\quad
\sin\gamma = \frac{\lambda_+-\lambda_-}{d}.
\end{equation}

With this convention, the range of $\gamma$ corresponding to Hale's law varies from cycle to cycle and between the two hemispheres. For cycle 25, in the northern hemisphere,  Hale-oriented BMRs are those with  $|\gamma| <90^\circ$. Anti-Hale regions have $|\gamma| >90^\circ$. In the northern hemisphere Joy's law is satisfied if either $\gamma>90^\circ$ or $-90^\circ<\gamma<0$ (independently of the cycle), and 
in the southern hemisphere if $\gamma<-90^\circ$ or
$0<\gamma<90^\circ$. Regions not satisfying Joy's law are called non-Joy regions. Please note that the reference to Joy’s law \citep{HE1919} here concerns the orientation of the bipolar region, and not the cited Joy’s law that often relates the increasing trend of the tilt angle with the emergence latitude.

\begin{figure}
    \includegraphics[width=\textwidth]{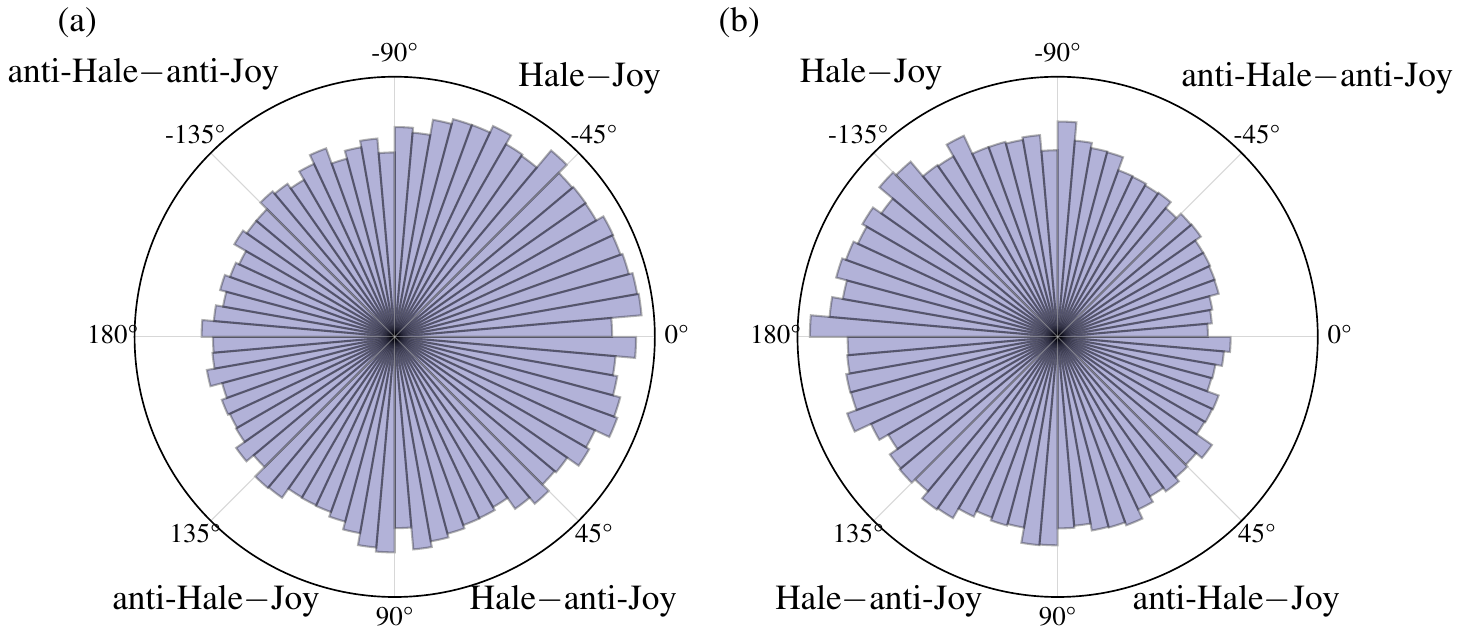}
    \caption{Circular histogram showing tilt distribution of tracked ERs. (a)  Northern Hemisphere. Here regions with $|\gamma|< 90^\circ$ follow Hale polarity rule, (b) Southern Hemisphere. Here tilt angles \anu{$|\gamma| >90^\circ$} are from regions following the Hale polarity rule.}
\label{fig:tilt_dist}
\end{figure}

BMRs predominantly obey Hale's polarity law, with only $\sim8$\% of regions violating Hale's polarity rule, and statistically exhibit an increase in tilt with latitude, Joy's law \citep{MN2014, Li2018, JN2021}. The tendency of ERs to obey these two laws is much weaker \citep{H2001}.
Figure~\ref{fig:tilt_dist} presents the circular histogram of ER tilt orientations for the northern and southern hemispheres separately. We observe a weak preference for Hale-oriented configurations: approximately 54\% of the ERs fall into the Hale category, consistent with the findings of \citet{TV2010} and earlier studies such as \citet{H2001} and \citet{TP2011}. 
Of the 54.2\% of Hale regions, 27.94\% exhibit a \anu{Hale--Joy orientation, while 26.30\% exhibit a Hale--anti-Joy orientation. Among the 45.8\% of anti-Hale regions, 23.25\% are Joy, whereas 22.51\% are classified as anti-Joy regions.} The mean tilt angle in Figure~\ref{fig:tilt_dist}(a) is $-7.28^\circ$, with a circular variance of $V=0.93$, while in Figure~\ref{fig:tilt_dist}(b), the mean tilt is $-178.34^\circ$, with $V=0.92$. These values, as suggested from the figure, indicate a weak preference for the Hale--Joy orientation, but with a large spread in the tilt distribution.
We see a much weaker tendency for the ERs to obey Joy's law, that is to have the leading polarity closer to the equator than the trailing polarity. Regions obeying Joys law
in the northern hemisphere lie in the upper right or lower left quadrant of Figure~\ref{fig:tilt_dist}(a). 
In the southern hemisphere, Figure~\ref{fig:tilt_dist}(b), those obeying Joy's law lie in the upper-left or lower-right quadrants. 

Although ERs in our study have a weak statistical tendency to obey both Hale's law and Joy's law, it is unclear how relevant this is for the large scale dynamo because we found the approximately same number of ERs with $\gamma<0$ and $\gamma>0$ in both hemispheres. The distinction in the sign of $\gamma$ is important because it determines how much net flux is transported between the hemispheres, and having the same number of regions with $\gamma>0$ as for $\gamma<0$ suggests little flux will be transported across the equator by ER emergence. We also examined whether the ERs show a tendency to follow Joy's law by testing whether the mean tilt increases with latitude. We do not find a significant increase in tilt with latitude.


\section{Conclusion and Discussion}
Unlike sunspots,
the small-scale magnetic bipoles, such as ERs, remain relatively less understood. These short-lived bipolar magnetic features, with fluxes on the order of $10^{20}$ Mx, have long posed challenges in their definition, identification, and statistical characterization, and their implications in the solar dynamo. Moreover, the lifetime and fluxes of ERs are consistent with the lifetimes observed of coronal bright points \citep{M2019, nina25}. Since their initial discovery in the 1970s, ERs have been described as a distinct population of short-lived bipolar magnetic features, spanning flux ranges of $\sim10^{20}$ Mx and lifetimes of a few hours to days, distributed ubiquitously across the solar surface. They were considered a separate class because, unlike large-scale flux emergence, they exhibited a highly disordered orientation and showed only a weak tendency to follow Hale's polarity rule. Early studies lacked spatial and temporal resolution to precisely trace their evolution, often leaving uncertainties about their physical origin and relationship to the global solar dynamo.

In this study, we revisit ERs by identifying and tracking them in the HMI LOS magnetograms. We analyze a continuous five-year dataset (January 2020 –- December 2024) of HMI LOS magnetograms to systematically detect and track ERs. To accomplish this, we developed a fully automated detection and tracking pipeline designed to identify small-scale bipolar features, associate opposite-polarity elements based on spatial proximity and flux balance, and follow their temporal evolution throughout their lifetime. The algorithm ensures that evolving phases of the ERs are consistently tracked across consecutive magnetograms, allowing us to build a statistically robust catalogue of  $\sim4\times10^{5}$ tracked ERs.

The detected ERs span a total unsigned flux range of $10^{18}-10^{20}$ Mx and exhibit a characteristic log-normal distribution. Notably, the distribution remains continuous between $10^{20}-10^{21}$Mx, traditionally regarded as the transition between ERs and strong active regions. This result indicates the view that solar magnetic flux on the Sun forms a continuous spectrum rather than falling into two distinctly separated populations. Large BMRs and small ERs therefore appear to represent different manifestations of the same underlying magnetic processes acting across multiple spatial scales.

Their spatial distribution shows only a weak dependence on latitude. Unlike sunspot emergence, which is strongly concentrated in mid-low latitudes, ERs occur across nearly the entire solar disk, with only a weak preference toward low-mid latitudes. Latitude-binned flux emergence rates likewise show a slight enhancement at these higher latitudes. However, over the course of the solar cycle, we observe a weak equatorward migration of the active latitude bands. Another important property of BMRs is the orientation of their emergence. We find that 54\% of the detected ERs adhere to Hale's polarity rule, which is considerably lower than the corresponding value for active regions, where approximately 90\% of regions obey Hale's polarity rule.

These results present a mixed picture. On one hand, we only observe a weak latitudinal confinement of ER emergence, in comparison with BMRs, which is a well-known characteristic of the large-scale dynamo. On the other hand, we do observe weak signatures of solar cycle properties in the detected ERs, indicating a weak linkage of ERs to the global dynamo. This suggests that the catalogue likely comprises a mixed population of ERs. A fraction of the detected regions probably originate from the large-scale dynamo, arising from the same flux emergence process that governs active regions, but representing the lower end of the magnetic flux emergence spectrum on the solar surface. This interpretation is supported by the continuous flux distribution obtained in our catalogue. Previous studies have shown that, at lower flux regimes, the latitudinal distribution of emerging regions becomes much broader than that of sunspots \citep{SJ2023}. Similarly, in our analysis, ERs emerge over a much wider range of latitudes than active regions. Owing to their relatively low magnetic flux, these regions are more susceptible to convective buffeting while rising through the convection zone, which can introduce a large dispersion in their tilt angles and weaken their adherence to Hale's and Joy's laws. The remaining fraction of the detected ERs likely originates from the small-scale turbulent dynamo, produced by the shredding and tangling of large-scale magnetic fields, whose properties are largely disconnected from the global dynamo. The weak adherence of the detected ERs to Hale's polarity rule further supports this interpretation of a mixed population.

In conclusion, our statistical analysis provides evidence that ERs represent a critical link in the continuum of magnetic flux emergence on the Sun. Their flux and spatial distributions, as well as their weak adherence to global polarity patterns, all point toward the identified ERs being a mixture of flux associated with the small-scale and 
large-scale (global) dynamos. 
Despite their modest flux content, these tiny BMRs continuously inject magnetic flux into the photosphere, contributing to the maintenance of the quiet-Sun magnetic network, and help in energy transport to upper atmospheres.

\section*{Acknowledgements}
B.B.K. acknowledges the support provided by the Alexander von Humboldt Foundation during the initial phase of this project. He further acknowledges the financial support from the Anusandhan National Research Foundation (ANRF) through the MATRIC program (file no. MTR/2023/000670) and Indian Space Research Organisation (project no. ISRO/RES/RAC-S/IITBHU/2024-25). The LOS magnetogram data used for the analyses in this article are obtained from MDI and HMI and downloaded via JSOC, courtesy of the NASA/SOHO and NASA/SOHO and SDO science teams.

\bibliographystyle{spr-mp-sola}
\bibliography{paper}


\label{lastpage}
\end{document}